\documentclass[twocolumn,preprintnumbers,amsmath,amssymb]{revtex4-1}
\usepackage[latin9]{inputenc}
\makeatletter
\newcommand{\rmd}{{\rm d}}

\newcommand{\fvec}[1]{\boldsymbol{#1}}

\usepackage{graphicx}
\usepackage{dcolumn}
\usepackage{bm}
\usepackage{epsfig}\usepackage{epstopdf}

\begin{document}

\title{Competing magnetic orders in the superconducting state of Nd-doped
CeRhIn$_{5}$ under pressure}

\author{P. F. S. Rosa$^{1}$, Jian Kang$^{3}$, Yongkang Luo$^{1}$, N. Wakeham$^{1}$,
E. D. Bauer$^{1}$, F. Ronning$^{1}$, Z. Fisk$^{2}$, R. M. Fernandes$^{3}$
and J. D. Thompson$^{1}$}

\affiliation{ $^{1}$ Los Alamos National Laboratory. Los Alamos, New Mexico 87545,
U.S.A.~\\
 $^{2}$ University of California, Irvine, California 92697-4574,
U.S.A.~\\
 $^{3}$ School of Physics and Astronomy, University of Minnesota,
Minneapolis, Minnesota 55455, USA}

\date{\today}
\begin{abstract}
Applied pressure drives the heavy-fermion antiferromagnet CeRhIn$_{5}$
towards a quantum critical point that becomes hidden by a dome of
unconventional superconductivity. Magnetic fields suppress this superconducting
dome, unveiling the quantum phase transition of local character.
 Here, we show that $5\%$ magnetic substitution at the Ce
site in CeRhIn$_{5}$, either by Nd or Gd, induces a zero-field magnetic
instability inside the superconducting state. This magnetic
state not only should have a different ordering vector than the high-field
local-moment magnetic state, but it also competes with the latter,
suggesting that a spin-density-wave phase is stabilized in zero field by Nd and Gd impurities
\textendash{} similarly to the case of Ce$_{0.95}$Nd$_{0.05}$CoIn$_{5}$.
Supported by model calculations, we attribute this spin-density wave
instability to a magnetic-impurity driven condensation of the spin
excitons that form inside the unconventional superconducting state.
\end{abstract}
\maketitle

Unconventional superconductivity (SC) frequently is found as an antiferromagnetic (AFM) transition is tuned by chemical substitution or pressure toward a zero-temperature phase transition, a 
magnetic quantum-critical point. This observation has a qualitative explanation: the proliferation of quantum fluctuations of magnetic origin at low temperatures can trigger the formation of a new ordered 
state. Unconventional superconductivity is a natural candidate state because it can be induced by an attractive Cooper-pair interaction provided by the fluctuating magnetism \cite{Monthoux, Scalapino}. Typical examples 
include copper-oxides, which without chemical substitution are AFM Mott insulators \cite{Cuprates}, metallic iron-based antiferromagnets that superconduct under pressure or with chemical substitutions \cite{FeAs}, 
and rare-earth heavy-fermion compounds with large effective electronic masses \cite{HF}.  

A characteristic manifestation of the unconventional nature of the
superconducting state is the momentum dependence of the SC
gap $\Delta$ that develops below the superconducting transition temperature
($T_{c}$). In contrast to conventional superconductors, $\Delta$ is not uniform
but instead has different signs in different regions of the Fermi
surface. Despite the distinct chemical and electronic 
properties of these materials, the interplay between magnetism and
SC is common among them, calling for a deep understanding
of this relationship. In this regard, heavy-fermion materials offer an ideal platform to
explore the relationship between these two phases.

An additional common feature among these different classes of superconductors is the emergence of a collective
magnetic excitation below $T_{c}$ often attributed to the formation of a spin exciton
\cite{Eschrig06}. This collective mode, whose energy has been shown
to scale with $\Delta$ across different materials \cite{NatPhysINS},
is a direct consequence of the sign-changing nature of $\Delta$. An example is the heavy-fermion superconductor CeCoIn$_{5}$,
known to be very close to an AFM quantum-critical point without tuning
\cite{HF}. Indeed, its SC gap is sign-changing \cite{WKPark,Zhou}
and a spin resonance mode is observed below $T_{c}$ \cite{stock}.
The energy of this mode in CeCoIn$_{5}$ scales
with its SC gap with the same proportionality found in copper-oxide
and iron-based systems.

Recent inelastic neutron scattering 
experiments find that the resonance mode in CeCoIn$_{5}$ is incommensurate at  the wavevector $\mathbf{Q}=(0.45,0.45,0.5)$  \cite{INS2015}. 
Due to Ce's $\it4f$ crystal-field environment, this mode is a doublet and the corresponding fluctuations are polarized along the $c$-axis . When a magnetic field $H$ is applied in the 
tetragonal $ab$-plane, this mode splits into two well-defined branches \cite{panarin,stock2012}. The field dependence of the Zeeman-split lower energy mode extrapolates to zero energy at 
$\sim 110$~kOe, which is remarkably close to the field where long-range AFM order develops inside the low-$T$, high-$H$ SC state \cite{UrbanoQ,KenzelmannScience, QPhase}. 
Spin-density wave (SDW) order in this so-called $Q$-phase has a small
$c$-axis ordered moment of $0.15$~$\mu_{B}$, which corresponds
closely to the spectral weight of the low-energy resonance mode. Moreover,
the SDW displays the same incommensurate wave-vector $\mathbf{Q}$ as the spin resonance
mode \cite{KenzelmannScience}. These observations suggest that the $Q$-phase is the result of
a condensation of spin excitations \cite{INS2015,stock2012,Mineev2011}.

In addition to the field-induced $Q$-phase, AFM order is found in Ce$_{0.95}$Nd$_{0.05}$CoIn$_{5}$ below $T_{c}$, in this case at zero field \cite{CeNdPetrovic}. The wave-vector and moment size of the 
Nd-induced magnetism are the same as those observed in the $Q$-phase of CeCoIn$_{5}$ \cite{CeNdNeutrons}. Although the sign-changing $\Delta$, with its nodes on the Fermi surface, plays a 
non-trivial role in enabling these orders, the obvious similarity between $H$- and Nd-induced magnetism strongly suggests that they have a common origin, namely 
condensation of the spin excitations that give rise to the resonance mode.

No evidence for the $Q$-phase has been found in other Ce$M$In$_{5}$ members ($M =$ Rh, Ir) or, for that matter, in any other superconductor. It is uncommon to find a magnetic 
transition below $T_{c}$ when both superconducting and magnetic states arise from the same electrons. Besides the example of CeCoIn$_{5}$, field-induced magnetism has been observed in La$_{1.9}$Sr$_{0.1}$CuO$_{4}$ \cite{Lake, Demler}. 
This AFM order, however, is distinct from a  $Q$-like 
 phase and is closely related to the field-induced magnetism in the SC state of pressurized CeRhIn$_{5}$ \cite{TusonNature}. At zero pressure, CeRhIn$_{5}$ displays AFM order at 
 $T_{N}=3.8$~K and $\mathbf{Q_{\mathrm{AFM}}}=(0.5,0.5,0.297)$ \cite{BaoCeRhIn5}. Pressurizing CeRhIn$_{5}$ tunes its magnetic transition toward a quantum-critical point and induces SC that coexists with AFM order for pressures up to $P_{c1} = 1.75$~GPa, where $T_{c}$ equals $T_{N}$.  Above $P_{c1}$, evidence for $T_{N}$ is 
 absent and only SC is observed \cite{TusonNature,Hegger, mito}. Application of a magnetic field, however, induces magnetism in the SC state between $P_{c1}$ 
and the quantum-critical point at $P_{c2} \sim 2.3$~GPa \cite{TusonNature, knebel}. Unlike magnetic order in the $Q$-phase, which exists only inside the SC state, field-induced magnetism in CeRhIn$_{5}$  persists into the normal state above  
the Pauli-limited $H_{c2}$ and is a smooth continuation of the zero-field $T_{N}(P)$ boundary \cite{TusonNature, knebel}.  This magnetism may obscure or preempt the formation of a  $Q$-like phase, but strong 
similarities of  CeCoIn$_{5}$ to CeRhIn$_{5}$ at $P>P_{c1}$ \cite{TusonNature, Pham} suggest the possibility that AFM order might develop in the high pressure SC state of Ce$_{1-x}$Nd$_{x}$RhIn$_{5}$ in 
zero field.

In this paper, we show that Nd induces a zero-field phase transition
in the high-pressure SC phase of Ce$_{0.95}$Nd$_{0.05}$RhIn$_{5}$
and present evidence that the phase transition is due to magnetic
order. This result generalizes the observation of magnetic order below
$T_{c}$ in Ce$_{0.95}$Nd$_{0.05}$CoIn$_{5}$ because pressure suppresses the 
magnetic order at the same rate in both compounds. Our model calculations support our conclusion that the magnetism in
Nd-doped CeRhIn$_{5}$ is due to the condensation of spin excitations
promoted by magnetic impurity scattering, and is thus distinct from
the local-moment magnetism in pure CeRhIn$_{5}$ promoted by the application of magnetic
fields. In agreement with this proposal, we
observe a competition between the field-induced magnetism, which displays
the same behavior as in CeRhIn$_{5}$, and the Nd-induced magnetism
in zero field. Hence, we expect a spin resonance with $c$-axis character below $T_{c}$ in CeRhIn$_{5}$ at pressures greater than  $P_{c1}$. More generally, our work reveals a route
to induce zero-field magnetic order via chemical substitution of magnetic
impurities in other unconventional superconductors that host spin
resonance modes.

\section{Results}

\begin{figure}[!hb]
\begin{center}
\includegraphics[width=0.9\columnwidth]{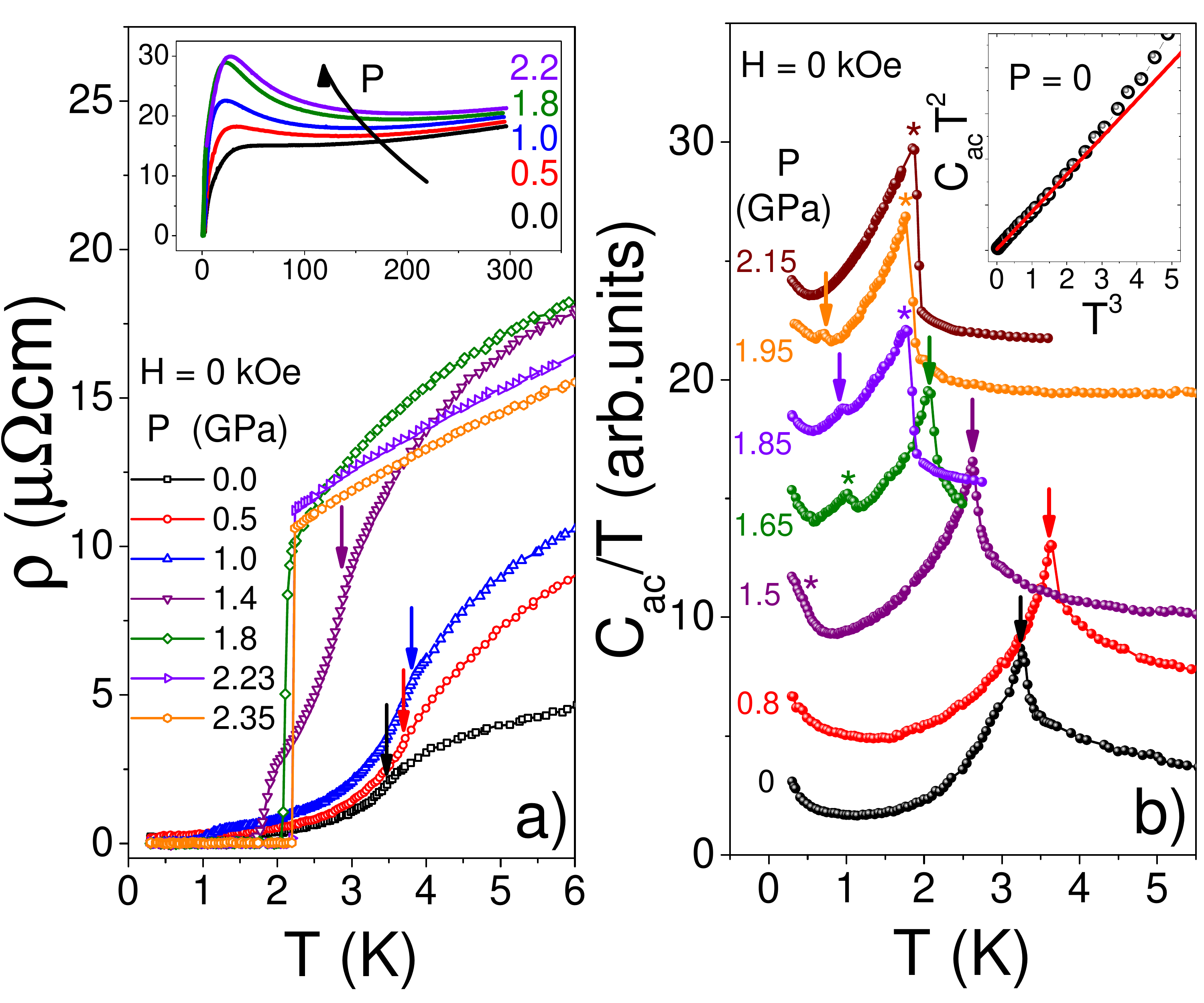}
\end{center}
\caption{a) Low-$T$ dependence of the in-plane electrical resistivity, $\rho(T)$, of Ce$_{0.95}$Nd$_{0.05}$RhIn$_{5}$ (s1) under pressure. Arrows mark $T_{N}$ determined by peaks in the first derivative. The inset shows $\rho(T)$ over the entire $T$-range. b) $C_{ac}/T$ $vs$ $T$ for Ce$_{0.95}$Nd$_{0.05}$RhIn$_{5}$ (s2) under pressure. A vertical offset of 2.5 units is 
added for clarity. Arrows (asterisks) denote $T_{N}$ ($T_{c}$). Inset shows a linear fit in a $C_{ac}T^{2}$ $vs$ $T^{3}$ 
plot.} 
\label{fig:Fig1}
\end{figure}

For comparison with Ce$_{0.95}$Nd$_{0.05}$CoIn$_{5}$, we grew crystals of 
Ce$_{0.95}$Nd$_{0.05}$RhIn$_{5}$ by an In-flux technique \cite{CeNdMe} and studied its pressure and 
field dependence by electrical resistivity and AC calorimetry measurements (See Methods for details). 
Figure~\ref{fig:Fig1}a shows the low-temperature  electrical resistivity, $\rho(T)$, on sample s1 at representative pressures, and the inset displays $\rho(T)$ in the whole $T$-range. Although Nd-substitution reduces $T_{N}(P=0)$ from 3.8 K to 3.4 K and slightly increases the residual resistivity $\rho_{0}$ to 0.2~$\mu\Omega.cm$, the $P$-dependence reported in Fig. 1a is essentially identical to that of CeRhIn$_{5}$ below $P^{*}_{c1}=1.8$~GPa where $T_{c}$ equals $T_{N}$. We also note that, at zero pressure, the $H-T$ phase diagram of Ce$_{0.95}$Nd$_{0.05}$RhIn$_{5}$ closely resembles the one found in CeRhIn$_{5}$.  These results indicate that $5\%$ Nd does not change drastically the local AFM character of $T_{N}$ below $P^{*}_{c1}$. Once $T_{c}$ exceeds $T_{N}$ at $P^{*}_{c1}$, there is no evidence for magnetism 
in $\rho(T)$, and the possibility of Nd-induced magnetism is obscured by the zero-resistance state below $T_{c}$. To investigate whether there is AFM order in the SC state, heat capacity measurements are necessary.

Figure~\ref{fig:Fig1}b shows the $T$-dependence of heat capacity divided by temperature ($C_{\mathrm{ac}}/T$) for sample s2 at various pressures. At 
ambient pressure, $C_{\mathrm{ac}}/T$  peaks at $T_{N}$ as in CeRhIn$_{5}$. As $T$ is lowered further, 
however, $C_{ac}/T$ turns up below $\sim 1$~K, which was not observed in CeRhIn$_{5}$ \cite{KHnote}. This upturn is presumably associated
with the nuclear moment of Nd ions, and it can be fit well by a sum of electronic ($
\propto \gamma$) and nuclear ($\propto T^{-3}$) terms \cite{SchottkyNuclear}.  The inset of Fig.~1b shows that
$C_{ac}T^{2}$ is linear in $T^{3}$, consistent with the presence of a nuclear Schottky contribution. We note, 
however, that an upturn also is observed at $2.3$~GPa (not shown) where magnetic order is absent, suggesting that the hyperfine 
field may not be solely responsible for splitting the nuclear levels. Although Nd nuclei have large zero-field quadrupole 
moments, Kondo-hole physics cannot be ruled out as a possible source of the upturn \cite{KHPagliuso,KHEric}.

For pressures below $P^{*}_{c1}$, $T_{N}$ evolves with $P$ as it does in transport data. Evidence for bulk SC (marked by asterisks), however, is observed at lower temperatures relative to the zero-resistance  state 
in $\rho(T)$. A difference between zero-resistance and bulk SC transitions also appears in CeRhIn$_{5}$ and has been attributed to filamentary SC due to the presence of long-range 
AFM order \cite{TusonNewJ}. Unlike CeRhIn$_{5}$ at pressures greater than $P_{c1}$, however, there is evidence for a phase transition in the SC state of Ce$_{0.95}$Nd$_{0.05}$RhIn$_{5}$ without an applied field. At $1.85$~GPa, an anomaly in 
$C_{ac}/T$ is observed at $1$~K (arrow in Fig.~\ref{fig:Fig1}b), below the SC transition at $T_{c}$~=~$1.77$~K. 
For reasons discussed below,
this anomaly stems from a magnetic order, and it is fundamentally
different from the AFM order
 displayed by the system for pressures smaller than $P_{c1}$.
The shape and magnitude of the anomaly relative to that at $T_{c}$ are very similar to those at $T_{N}$ in Ce$_{0.95}$Nd$_{0.05}$CoIn$_{5}$ (see Supplemental Materials), and the small entropy associated with it suggests that the magnetic order is most likely a small-moment density wave. As we will come to later, this evidence is most obvious in data shown in Fig.~\ref{fig:Fig3}.

\begin{figure}[!ht]
\begin{center}
\includegraphics[width=0.8\columnwidth,keepaspectratio]{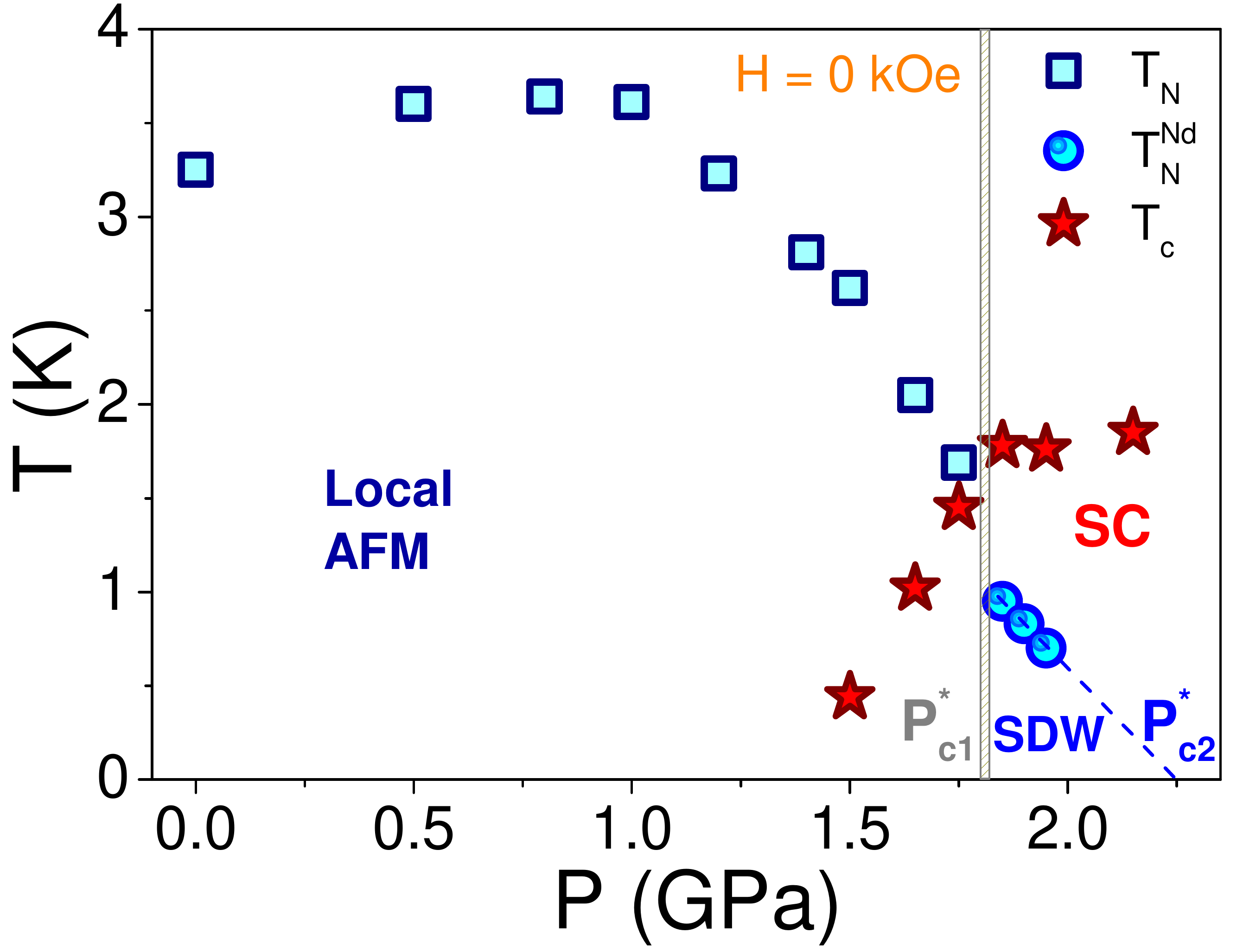}
\end{center}
\caption{Zero-field $T-P$ phase diagram of
 Ce$_{0.95}$Nd$_{0.05}$RhIn$_{5}$ (s2) obtained from AC calorimetry measurements. Here $P^{*}_{c1}\sim1.8$~GPa and $P^{*}_{c2} \sim 2.3$~GPa.}
\label{fig:Fig2}
\end{figure}

\begin{figure*}[!ht]
\begin{center}
\includegraphics[width=1.4\columnwidth,keepaspectratio]{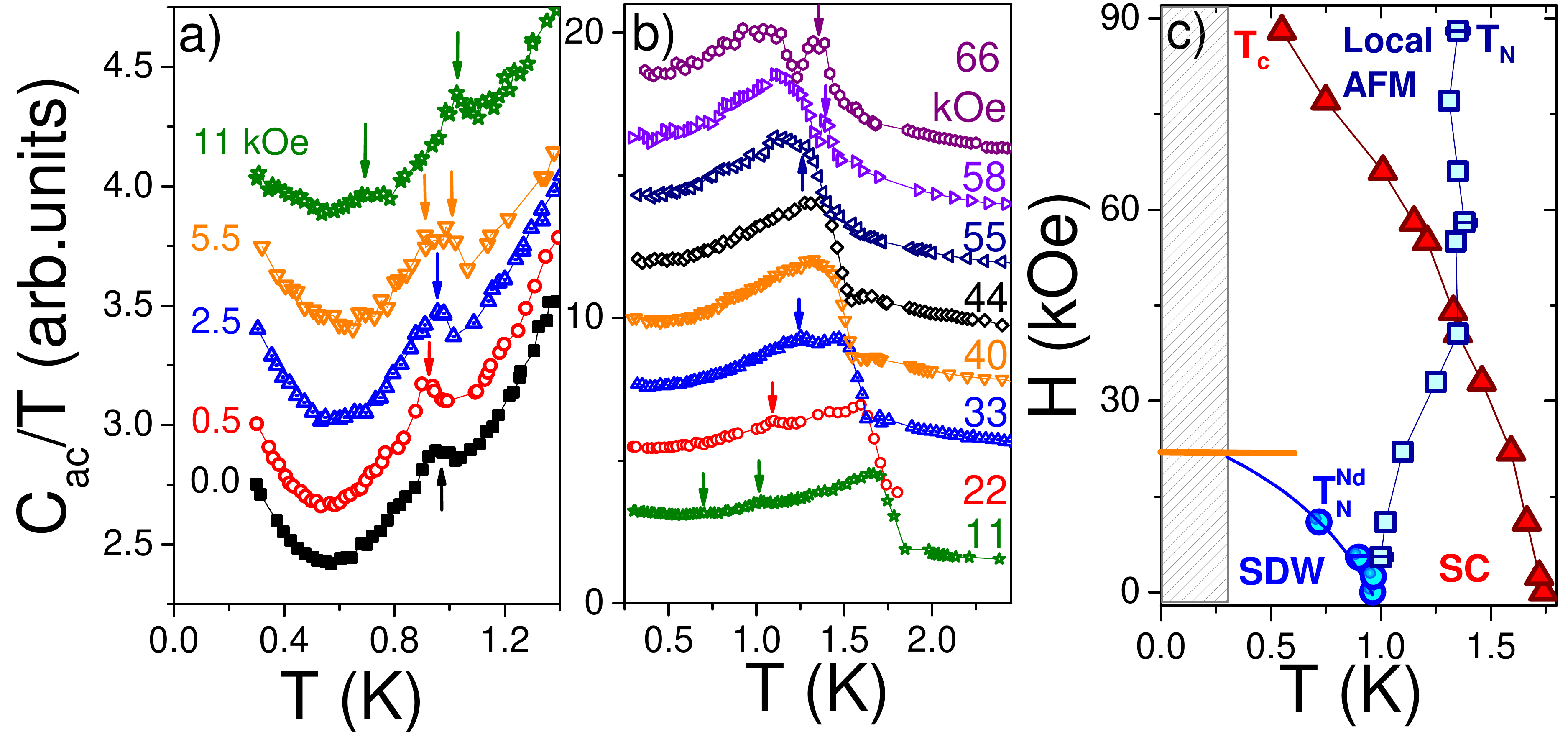}
\end{center}
\caption{AC heat capacity, $C_{ac}$, of Ce$_{0.95}$Nd$_{0.05}$RhIn$_{5}$ (s2) at 1.85~GPa. a)  $T$-dependence of $C_{ac}/T$ at low fields. An offset of 0.2 has been added for clarity. b) $T$-dependence of $C_{ac}/T$ at high fields. c) $H-T$ phase diagram. The diagonal bars delimit the inaccessible temperature region in our experiments ($T<0.3$~K). The solid horizontal line at $H=22$~kOe indicates that no transition is observed above $0.3$~K for this field.}
\label{fig:Fig3}
\end{figure*}

We summarize the zero-field results discussed above in the $T$-$P$ phase diagram shown in Fig.~\ref{fig:Fig2}. Local-moment-like 
AFM order coexists with bulk SC in a narrow pressure range 
below $P^{*}_{c1}$. From just below to just above $P^{*}_{c1}$, $T_{N}(P)$ changes discontinuously, in contrast to field-induced magnetism 
in CeRhIn$_{5}$, which is a smooth continuation of $T_{N}(P)$ from below $P_{c1}$ \cite{TusonNature}. This supports the interpretation that $H$- and Nd-induced magnetic orders have different 
origins. Therefore, the Nd-induced transition is labeled $T^{\mathrm{Nd}}_{N}$ to distinguish it from 
$T_{N}$ in pure CeRhIn$_{5}$. Above $P^{*}_{c1}$, $T^{\mathrm{Nd}}_{N}$
 is suppressed at a rate of $-2.4$~K/GPa and extrapolates to zero temperature, i.e. a quantum-critical point, at 
 $P^{*}_{c2} \sim 2.3$ GPa  inside the superconducting phase. Whether the coincidence of $P^{*}_{c2}$ and  $P_{c2}$ is accidental or not requires further
  investigation beyond the scope of our work. As shown in the Supplemental Materials, the
 rate of suppression of this Nd-induced transition is the same rate found in 
Ce$_{0.95}$Nd$_{0.05}$CoIn$_{5}$ within experimental uncertainty, strongly indicating a common origin.  Because entropy associated 
with the zero-field transition is rather small, as found in Ce$_{1-x}$Nd$_{x}$CoIn$_{5}$, the typical signature of 
quantum criticality (i.e., divergence of $C/T$ at  low-$T$) is likely hidden by SC 
and by the upturn in $C/T$. We also note that the highest $T_{c}$ achieved in 
Ce$_{0.95}$Nd$_{0.05}$RhIn$_{5}$, 
$T^{\mathrm{max}}_{c}=1.85$~K, is $0.4$~K lower than $T^{\mathrm{max}}_{c}$ of CeRhIn$_{5}$. This same 
suppression of $T^{\mathrm{max}}_{c}$ is observed in Ce$_{0.95}$Nd$_{0.05}$CoIn$_{5}$ and 
indicates that Nd ions act similarly as magnetic pair-breaking impurities. 

To further investigate the nature of the Nd-induced magnetism, we turn to the field-dependent heat capacity data. 
Figure~\ref{fig:Fig3}a shows $C_{\mathrm{ac}}/T$ $vs$ $T$ at $1.85$~GPa ($>~P_{c1}$) and low 
magnetic fields. The zero-field transition at $T^{\mathrm{Nd}}_{N}~=~0.96$~K remains unchanged in a field of $2.5$~kOe.
  As the field is increased further, however, the specific heat anomaly splits, one anomaly moving to lower temperatures 
and the other to higher temperatures. At $11$~kOe, our data show
features at $1$~K and $0.7$~K. Previous reports on CeRhIn$_{5}$  at similar pressures ($\sim 1.9$~GPa) show that the
 field-induced transition emerges at  $T_{N}=1$~K when $H=11$~kOe 
\cite{TusonNewJ}. It is thus reasonable to associate the anomaly we observe at $1$~K and 11~kOe with the field-induced 
magnetism in CeRhIn$_{5}$.  Further, Fig.~\ref{fig:Fig3}b shows the high-$H$
evolution of $T_{N}$ with a field dependence very similar to that of CeRhIn$_{5}$: $T_{N}$ first increases with $H$ and then 
remains constant above the upper critical field $H_{c2}$. As shown in the $H-T$ phase diagram (Fig.~3c), this field-induced $T_{N}$ clearly competes with 
$T^{\mathrm{Nd}}_{N}$, as reflected in the rapid suppression of $T^{\mathrm{Nd}}_{N}$ as a function of 
$H$. In fact, no evidence for $T^{\mathrm{Nd}}_{N}$ is 
observed at fields $H \geq 22\,$~kOe, implying a field-induced quantum-critical point in addition to the pressure-induced, zero-field critical point of this order. 
Due to the reasons explained above, our results strongly
point to two distinct types of magnetism emerging in 
Ce$_{0.95}$Nd$_{0.05}$RhIn$_{5}$. The first one is due to Nd ions and it has the hallmarks of that in 
Ce$_{0.95}$Nd$_{0.05}$CoIn$_{5}$. The second is the $H$-induced magnetism that appears in pure
CeRhIn$_{5}$ \cite{TusonNature,knebel}. 

\section{Discussion}

What is the role of Nd and why is it special? At a concentration of $5\%$, average spacing of $17\,$\AA$\,$ and 
non-periodic distribution on Ce-sites, Nd is too dilute to induce magnetic order by dipole 
 or indirect Ruderman-Kittel-Kasuya-Yosida interactions. Its role, then, must be more subtle. Using the bulk modulus of 
CeRhIn$_{5}$ and the unit-cell volume variation in 
Ce$_{1-x}$Nd$_{x}$RhIn$_{5}$, we estimate that Ce$_{0.95}$Nd$_{0.05}$RhIn$_{5}$ experiences an effective chemical pressure of
$\Delta P~=~0.25$~GPa relative to CeRhIn$_{5}$  \cite{CeNdMe}. From the phase diagram of CeRhIn$_{5}$, this $\Delta P$ would 
increase $T_{N}$ by  $ 0.1$~K instead of producing the observed reduction. 
Hence, we conclude that chemical pressure $\it per se$ is not the dominant tuning parameter. 

The disruption of translational periodicity 
of the Ce lattice by Nd substitution creates a ``Kondo hole" that contributes to reducing $T_{N}$ at zero pressure \cite{CeNdMe}. 
The latter conclusion is supported by the observation that non-magnetic La substitution for Ce in CeRhIn$_{5}$ also depresses $T_{N}$ similarly \cite{KHPagliuso,LeticiePRL}.  
 Neodymium, however, carries an additional magnetic moment that is unlikely to
be quenched by Kondo screening. In the context of CeCoIn$_{5}$,
Michal and Mineev \cite{Mineev2011} proposed that the $Q$-phase
observed in the presence of an in-plane magnetic field is the consequence
of the condensation of the spin-exciton collective mode found in the
SC phase. Thus, it is natural to consider whether the Nd magnetic
moments immersed in CeRhIn$_{5}$ at zero field could also promote a
similar behavior.

As discussed in detail in the Supplemental Materials,
condensation of spin excitons takes place when the spin-resonance-mode
frequency $\omega_{\mathrm{res}}$ vanishes. Within an random phase approximation (RPA) approach,
the latter is given by the pole of the renormalized magnetic susceptibility, i.e. when 
$\chi_{\mathrm{AFM}}\left(\mathbf{Q},\omega_{\mathrm{res}}\right)=1/U$,
where $U$ is the effective electronic interaction projected in the
SDW channel and $\chi_{\mathrm{AFM}}\left(\mathbf{Q},\omega_{\mathrm{res}}\right)$
is the non-interacting magnetic susceptibility inside the SC state.
When the ordering vector $\mathbf{Q}$ connects points of the Fermi
surface with different signs of the SC gap, \textbf{$\Delta_{\mathbf{k}}=-\Delta_{\mathbf{k}+\mathbf{Q}}$},
$\chi_{\mathrm{AFM}}\left(\mathbf{Q},\omega\right)$ generically diverges
when $\omega\rightarrow2\Delta$ and remains non-zero when $\omega\rightarrow0$.
Thus, even a very weak $U$ can in principle induce a
resonance mode with frequency near $2\Delta$. Once the interaction
increases, $\omega_{\mathrm{res}}$ moves to lower frequencies. When the
interaction overcomes a critical value, $U>U_{c}\equiv\chi_{\mathrm{AFM}}^{-1}\left(\mathbf{Q},0\right)$,
the resonance mode vanishes and SDW order is established inside the SC dome.

\begin{figure}[!ht]
\begin{center}
\includegraphics[width=0.9\columnwidth,keepaspectratio]{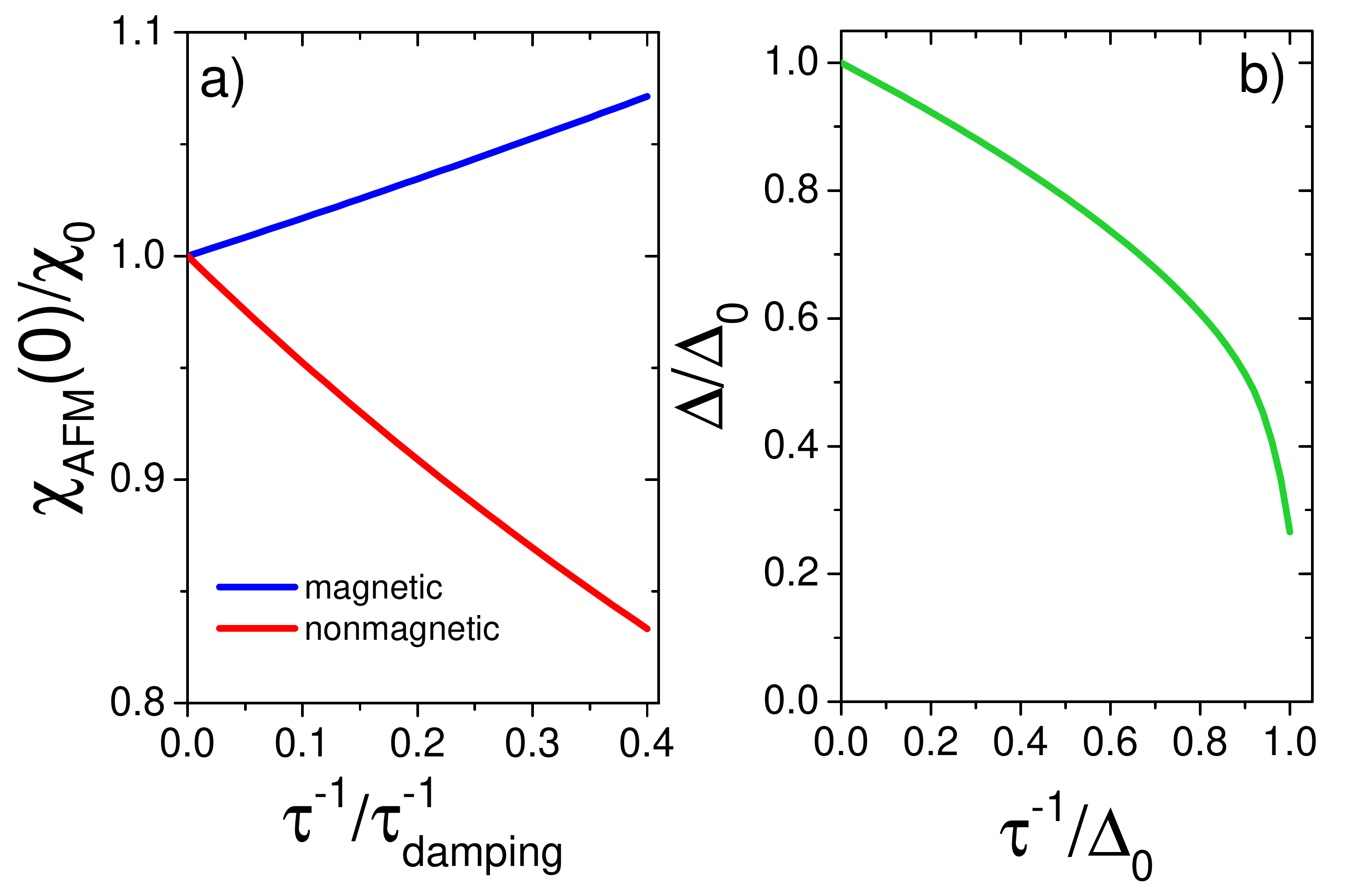}
\end{center}
\caption{a) Static spin susceptibility $\chi_{\mathrm{AFM}}\left(\mathbf{Q},\,\omega=0\right)$
as function of the total impurity scattering rate for the cases of
non-magnetic (red curve) and paramagnetic impurities (blue curve).
b) The suppression of the effective SC gap $\Delta$ by both magnetic
and non-magnetic impurities. In these plots, we considered point-like
impurities. Here, $\Delta_{0}$ is the gap of the clean system whereas
$\tau_{\mathrm{damping}}^{-1}$ is the Landau damping (see Supplemental
Materials for details).}
\label{fig:Fig4}
\end{figure}

In our case, the interaction $U$ is presumably independent of pressure.
Thus, in order for magnetic impurities to promote spin-exciton condensation,
$\chi_{\mathrm{AFM}}\left(\mathbf{Q},0\right)$ must increase (i.e.
the critical interaction value $U_{c}$ must decrease) as function
of the impurity potential. To investigate whether this is a sensible
scenario, we considered a toy model consisting of two ``hot spots''
located at momenta $\mathbf{k}$ and $\mathbf{k}+\mathbf{Q}$ at the
Fermi surface such that \textbf{$\Delta_{\mathbf{k}}=-\Delta_{\mathbf{k}+\mathbf{Q}}$}.
Note that such a hot-spots model has been previously employed to study
the effects of disorder on SC \cite{Kang16}. To focus on the general
properties of the model, we linearize the band dispersion around the
hot spots and compute both $\chi_{\mathrm{AFM}}\left(\mathbf{Q},0\right)$
and the effective gap amplitude $\Delta$ at $T=0$ as function
of the total impurity scattering rate $\tau^{-1}$ within the self-consistent
Born approximation (similarly to what was done in Ref. \cite{Chubukov11}
for $s^{+-}$ SC and perfectly nested bands). As shown in Figure 4,
whereas both magnetic and non-magnetic impurity scattering suppress
$\Delta$ at the same rate (panel 4b), we find that $\chi_{\mathrm{AFM}}\left(\mathbf{Q},0\right)$
is suppressed for non-magnetic impurity but enhanced by magnetic impurity
scattering (panel 4a). Thus, in the case of non-magnetic impurities, athough the
resonance mode frequency may decrease as compared to the clean case,
it never collapses to zero. Because magnetic impurity
scattering enhances $\chi_{\mathrm{AFM}}\left(\mathbf{Q},0\right)$
but not necessarily destroys SC, the system may undergo an SDW transition
inside the SC dome. Although the fate of the system will depend on microscopic
details beyond those captured by the toy model considered here, our model
nicely illustrates that it is plausible for magnetic impurity scattering
to drive spin-exciton condensation in the SC phase. In this regard,
we note that previous investigations of a microscopically-motivated
theoretical model also found that the $Q$-phase may be stabilized
by magnetic impurities even at zero external field \cite{Gastiasoro15}.

These results suggest that other magnetic impurities could induce
the same type of  SDW order in both CeCoIn$_{5}$ and
pressurized CeRhIn$_{5}$. In fact, we show in the Supplemental Materials
that easy-plane Gd$^{3+}$ ions ($J=S=7/2$) also induce a transition
in the heat capacity data of both Co and Rh members. This anomaly
is similar to the one induced by easy-axis ($c$-axis) Nd$^{3+}$ ions ($J=9/2$,
$L=3$, $S=3/2$) discussed above. Hence, Nd is not ``special'' in inducing magnetic
order in the superconducting states of Ce$_{0.95}$Nd$_{0.05}$CoIn$_{5}$
and Ce$_{0.95}$Nd$_{0.05}$RhIn$_{5}$, and other unconventional
superconductors that host a spin resonance mode may also display zero-field
magnetism via the same mechanism. Our results also imply
that non-magnetic impurities will not induce condensation of excitations, in agreement with experimental data on La-substituted CeRhIn$_{5}$
\cite{LeticiePRL,NeutronsLa,PDLaCo}.

Finally, we note that the SDW ordering
vector in Nd-doped CeCoIn$_{5}$ corresponds closely to the nodal structure \cite{INS2015}
 that is also found in the superconducting state of CeRhIn$_{5}$
\cite{park2008}. Hence, the magnetic wave-vector of zero-field order
above $P^{*}_{c1}$ in Ce$_{0.95}$Nd$_{0.05}$RhIn$_{5}$ should also
be close to \textbf{Q}$= (0.45,0.45,0.5)$, and we expect neutron scattering experiments
to find a spin resonance of $c$-axis character below $T_{c}$ in
CeRhIn$_{5}$ at $P>P_{c1}$.

\section{Conclusion}
In summary, we generalize the observation of Nd-induced magnetism
in CeCoIn$_{5}$ to pressurized CeRhIn$_{5}$ and Gd-substituted members.
This spin-density wave order, which reflects the nodal-gap symmetry,
 is argued to be a consequence of the condensation of spin excitations
that arise inside the SC state. Given the several similarities between
CeCoIn$_{5}$ and Ce$_{2}$PdIn$_{8}$ \cite{dong}, Nd substitution
might nucleate AFM order in its superconducting state. Appropriate
substitutions in other unconventional superconductors that host a
spin resonance also should induce zero-field magnetism by the same
mechanism, and magnetism should be tunable to a quantum-critical point
inside their SC phase.

We thank A. V. Chubukov, H. L$\mathrm{\ddot{o}}$hneysen, S. Maiti,
and P. G. Pagliuso for useful discussions. Work at Los Alamos by Y.L., N.W., E.D.B., F.R., and J.D.T. was
performed under the auspices of the U.S. Department of Energy, Office
of Basic Energy Sciences, Division of Materials Science and Engineering.
P. F. S. R. acknowledges a Director's Postdoctoral Fellowship through
the LANL LDRD program. The theoretical work (J.K. and R.M.F.) was
supported by the U.S. Department of Energy, Office of Science, Basic
Energy Sciences, under Award number DE-SC0012336.

\subsection*{Methods}
The series Ce$_{1-x}$Nd$_{x}$RhIn$_{5}$ was grown by the In-flux technique and its properties are reported elsewhere 
\cite{CeNdMe}. Crystals with \textit{x}=0.05 and free of unreacted In were 
mounted in a hybrid piston-cylinder pressure cell, filled with silicone fluid as the pressure medium, and a piece of Pb whose change in 
\textit{T}$_{c}$ served as a manometer. Gd-substituted crystals were grown using the same method. Electrical resistivity was measured by a four-probe method with current flow in the \textit{ab}-plane. 
Semi-quantitative heat capacity was obtained by an AC calorimetry technique described elsewhere \cite{sidorov}. Magnetic fields to 9 T were applied parallel to the \textit{ab}-plane. The results above have been reproduced in different crystals and, for clarity, we show resistivity and calorimetry data for two representative samples labeled sample 1 (s1) and sample 2 (s2), respectively.

\begin{center}

\newpage
\setcounter{table}{0}
\setcounter{figure}{0}
\onecolumngrid
\newpage

{\bf \huge
{\it Supplemental Materials}\\
\vskip 0.3cm
\Large Competing magnetic orders in the superconducting state of heavy-fermion CeRhIn$_{5}$}
\end{center}

\begin{center}
\author{P. F. S. Rosa$^{1}$, Jian Kang$^{3}$, Yongkang Luo$^{1}$, N. Wakeham$^{1}$,
E. D. Bauer$^{1}$, F. Ronning$^{1}$, Z. Fisk$^{2}$, R. M. Fernandes$^{3}$
and J. D. Thompson$^{1}$}

\affiliation{ $^{1}$ Los Alamos National Laboratory. Los Alamos, New Mexico 87545,
U.S.A.~\\
 $^{2}$ University of California, Irvine, California 92697-4574,
U.S.A.~\\
 $^{3}$ School of Physics and Astronomy, University of Minnesota,
Minneapolis, Minnesota 55455, USA}

\end{center}

\renewcommand{\thefigure}{S\arabic{figure}}
\renewcommand{\thetable}{S\arabic{table}}

\section{Supporting AC calorimetry measurements in CeCoIn$_{5}$ and CeRhIn$_{5}$}
\vskip 0.25cm

Single crystals of nominally Ce$_{0.95}$Nd$_{0.05}$CoIn$_{5}$  were grown from an excess In flux and characterized by 
pressure-dependent AC calorimetry, as described in the main text.  These measurements were carried out in a pressure cell under conditions 
also discussed in the main text. Figure~S1 shows $C_{\mathrm{ac}}/T$ as a function of $T$ at two pressures. At 
atmospheric pressure (top curve), a pronounced peak in $C_{\mathrm{ac}}/T$ defines the superconducting transition temperature 
$T_{c}$~=1.63 K that is followed at lower temperatures by an anomaly  peaked  near 0.7 K.  These features in data at atmospheric 
pressure coincide well with those reported in ref. [15] (main text) and indicate that the nominal Nd content is very close to the 
actual content.  Consequently, we associate the lower temperature anomaly with antiferromagnetic order at $T^{\mathrm{Nd}}_{N}$. We note 
that the shapes of the two anomalies resemble those found in Ce$_{0.95}$Nd$_{0.05}$RhIn$_{5}$.
As shown in Fig.~S1a, an applied pressure of only 0.13 GPa reduces  $T^{\mathrm{Nd}}_{N}$ to $\sim 0.4$~K. Assuming that $T^{\mathrm{Nd}}_{N}$  
decreases linearly with $P$ (Fig.~S1a inset) gives d$T^{\mathrm{Nd}}_{N}$/dP $\sim -2.3$~K/GPa. This rate of decrease is very close to that 
found in Fig. 2 of the main text where, for  Ce$_{0.95}$Nd$_{0.05}$RhIn$_{5}$ above $P_{c1}$,
 d$T^{\mathrm{Nd}}_{N}/$dP $\sim -2.4$~K/GPa.  The fact that Nd-induced zero-field antiferromagnetic order in similarly doped CeCoIn$_{5}$
   and CeRhIn$_{5}$  displays the same 
sensitivity to pressure strongly suggests that the zero-field magnetism has a common origin in both materials.  

\begin{figure}[!ht]
\begin{center}
\includegraphics[width=1\columnwidth]{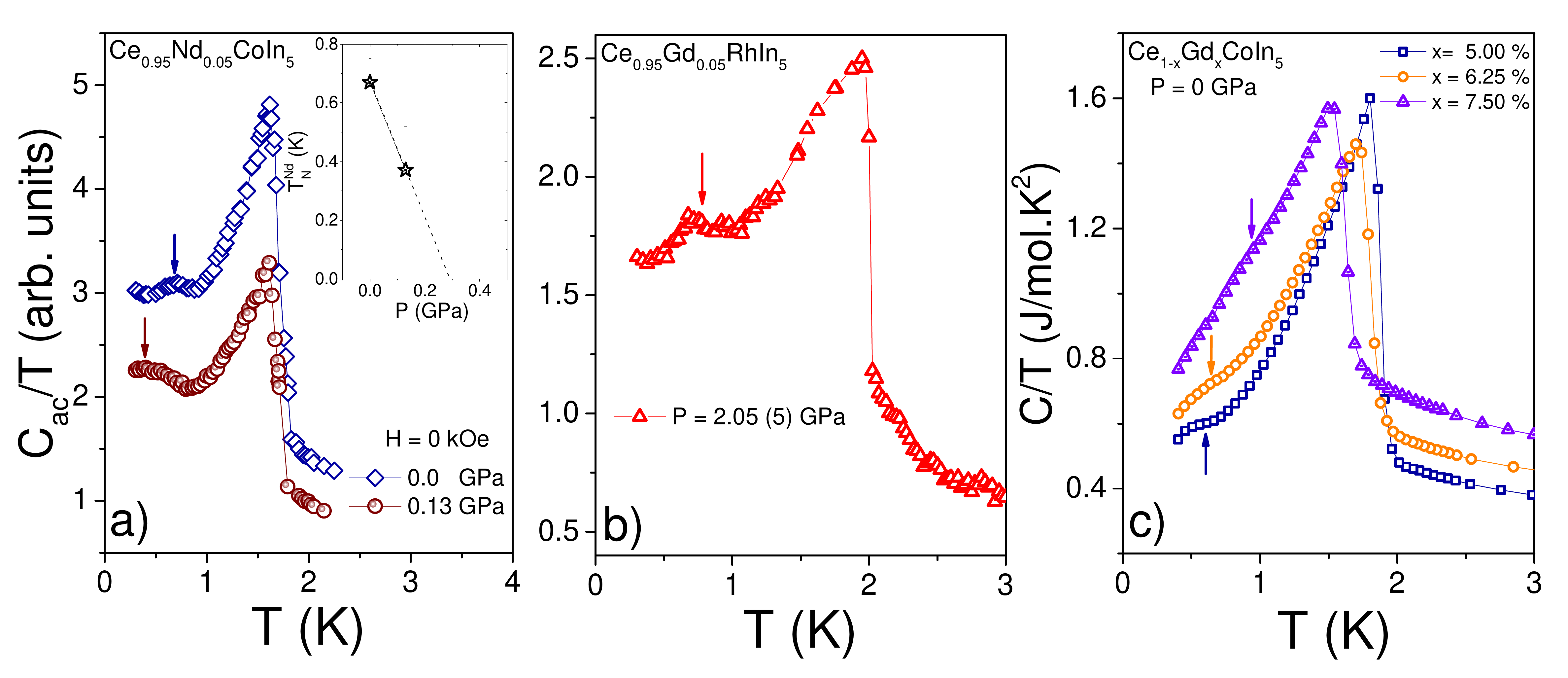}
\end{center}
\caption{(a) $C_{ac}/T$ $vs$ $T$ of Ce$_{0.95}$Nd$_{0.05}$CoIn$_{5}$ at two pressures. Arrows show $T_{N}$ and inset shows its extrapolation to $T=0$. (b) 
$C_{ac}/T$ $vs$ $T$ of Ce$_{0.95}$Gd$_{0.05}$CoIn$_{5}$ at 2.05~GPa. (c) $C_{ac}/T$ $vs$ $T$ of Ce$_{1-x}$Gd$_{x}$CoIn$_{5}$ ($x= 0.05, 0.0625, 0.075$) at ambient pressure. The curves are shifted by $0.1$ J/mol.K$^{2}$ for clarity. Arrows show $T_{N}$ determined as the minimum in the first derivative of the data.
}
\label{fig:Fig1SI}
\end{figure}

Single crystals of nominally Ce$_{0.95}$Gd$_{0.05}$RhIn$_{5}$  were grown from an excess In flux and characterized by 
pressure-dependent AC calorimetry, as described in the main text. Figure~S1b shows $C_{\mathrm{ac}}/T$ as a function of $T$ at 2.05~GPa.  A pronounced peak in $C_{\mathrm{ac}}/T$ defines the superconducting transition temperature 
$T_{c}$~=1.95 K that is followed at lower temperatures by an anomaly  peaked  near 0.7 K determined by the minimum in derivative of $C/T$.  

Finally, single crystals of nominally Ce$_{1-x}$Gd$_{x}$CoIn$_{5}$ ($x= 0.05, 0.0625, 0.075$)  were grown from an excess In flux and characterized by specific heat at ambient pressure in a commercial PPMS. Figure~S1c shows $C_{\mathrm{ac}}/T$ as a function of $T$ for all concentrations.  A pronounced peak in $C/T$ defines the superconducting transition temperature 
$T_{c}$~= $1.8$~K, $1.7$~K, and $1.52$~K that is followed at lower temperatures by an anomaly peaked near $0.6$~K, $0.7$~K, and $0.97$~K.

\newpage

\section{Impact of disorder on the magnetic susceptibility inside the superconducting
state}
\vskip 0.25cm

To gain insight into the problem, and keep the calculation analytically
tractable, we employ the hot spots approximation. In particular, we
consider two points on the Fermi surface, labeled $c$ and $d$, separared
by the SDW vector $\fvec Q$, such that $\Delta_{\fvec k}=-\Delta_{\fvec k+\fvec Q}$.
Next, we linearize the dispersions around the hot spots 
\begin{equation}
\epsilon_{c}(\fvec k)=\fvec v_{c}\cdot\fvec k\ ,\qquad\epsilon_{d}(\fvec k+\fvec Q)=\fvec v_{d}\cdot\fvec k
\end{equation}
where the momentum $\fvec k$ is measured with respect to $\fvec k_{F}$
and $\fvec v_{c/f}$ are the Fermi velocities. Here, we consider $v_{c}=v_{d}$
and let the relative angle between them be arbitrary but non-zero.
The case $\fvec v_{c}=-\fvec v_{d}$ corresponds to perfect nesting,
which was studied in Ref.~\cite{Chubukov11}. Here, we focus on the
case where nesting is not perfect. As long as the SDW instability
is driven by the low-energy electronic states, the linearized approximation
can be used to compute the leading contribution to the magnetic susceptibility.

There are two different types of impurity potentials: the non-magnetic
potential $u_{\fvec k}$, which couples to the charge degrees of freedom,
and the paramagnetic potential $\mathbf{u}_{\fvec k}^{p}$, which
couples to the spin degrees of freedom. Each impurity potential can
be split into small-momentum scattering, $u_{0}$ and $\mathbf{u}_{0}^{p}$
(which does not couple the fermions on the two hot spots), and large-momentum
scattering $u_{\fvec Q}$ and $\mathbf{u}_{\fvec Q}^{p}$ (which couples
the fermions on the two hot spots).

First, we study the impact of the various types of disorder on the
pairing gap. Here, we assume the existence of a SC state, without
discussing its origin. Thus, we consider a static pairing interaction
$V$ corresponding to a repulsive interaction coupling fermions from
different hot spots, such that $\Delta_{c}=-\Delta_{d}$. To proceed,
it is convenient to define Nambu spinors $\Psi_{i\fvec k}^{\dagger}=\left(f_{i,\fvec k\uparrow}^{\dagger},\,f_{i,-\fvec k\downarrow}\right)$.
The Green's function in Nambu space is then given by: 
\begin{equation}
G_{c}^{-1}=Z_{\omega}^{-1}(i\omega_{n}\sigma_{0}-\bar{\Delta}_{\omega}\sigma_{1})-\epsilon_{c}(\fvec k)\sigma_{3}\ ,\qquad G_{d}^{-1}=Z_{\omega}^{-1}(i\omega_{n}\sigma_{0}+\bar{\Delta}_{\omega}\sigma_{1})-\epsilon_{d}(\fvec k)\sigma_{3}
\end{equation}

\begin{figure}[htbp]
\begin{centering}
\includegraphics[width=0.9\columnwidth]{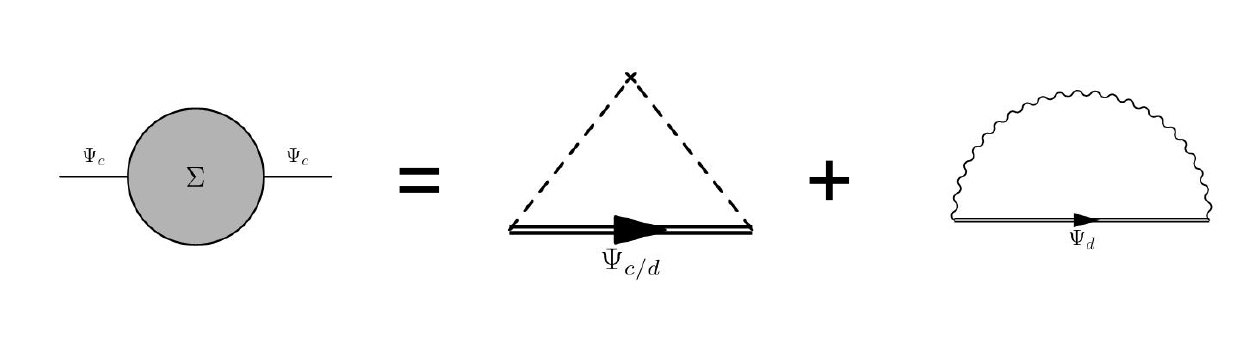}
\par\end{centering}
{\LARGE{}\caption{The fermionic self energy in the self-consistent approximation. The
two diagrams show the contributions arising from impurities and from
the pairing interaction, respectively.}
\label{Fig:FermionSE} }{\LARGE \par}
\end{figure}
where $Z_{\omega}$ is the imaginary part of the normal component
of the self-energy and $\bar{\Delta}_{\omega}$ (the gap normalized
by impurities) is the real part of the anomalous component of the
self-energy. As shown in Fig~\ref{Fig:FermionSE}, the fermionic
self-energy in the self-consistent approximation becomes, at $T=0$
\begin{align}
\Sigma_{c}(i\omega_{n})= & -\pi\big(u_{0}^{2}+u_{\fvec Q}^{2}+(u_{0}^{p})^{2}+(u_{\fvec Q}^{p})^{2}\big)N_{f}\frac{i\omega_{n}}{\sqrt{\omega_{n}^{2}+\bar{\Delta}_{\omega}^{2}}}\sigma_{0}+\pi\big(u_{0}^{2}-u_{\fvec Q}^{2}-(u_{0}^{p})^{2}+(u_{\fvec Q}^{p})^{2}\big)N_{f}\frac{\bar{\Delta}_{\omega}}{\sqrt{\omega_{n}^{2}+\bar{\Delta}_{\omega}^{2}}}\sigma_{1}\nonumber \\
 & +VN_{f}\int_{0}^{\Lambda}\frac{\bar{\Delta}_{\omega}\rmd\omega}{\sqrt{\omega^{2}+\bar{\Delta}_{\omega}}}\sigma_{1}
\end{align}

To solve for $Z_{\omega}$ and $\bar{\Delta}_{\omega}$, we average
over impurities and introduce the scattering rates $\tau_{0/\fvec Q}^{-1}=2\pi N_{f}u_{0/\fvec Q}^{2}$
for non-magnetic impurities and $\left(\tau_{0/\fvec Q}^{p}\right)^{-1}=2\pi N_{f}\sum_{i}\left(\mathbf{u}_{0/\fvec Q}^{p}\cdot\hat{\fvec e}_{i}\right)^{2}$
for paramagnetic impurities. In the hot-spots model, the density of
states is $N_{f}=\Lambda_{\parallel}/(2\pi)^{2}v_{F}$, with $\Lambda_{\parallel}$
denoting the momentum cutoff parallel to the Fermi surface. We obtain:
\begin{align}
Z^{-1}(\omega)= & 1+\left(\frac{1}{2\tau_{0}}+\frac{1}{2\tau_{\fvec Q}}+\frac{1}{2\tau_{0}^{p}}+\frac{1}{2\tau_{\fvec Q}^{p}}\right)\frac{1}{\sqrt{\bar{\Delta}_{\omega}^{2}+\omega^{2}}}\\
\bar{\Delta}_{\omega}= & -\frac{\bar{\Delta}_{\omega}}{\sqrt{\omega^{2}+\bar{\Delta}_{\omega}^{2}}}\left(\frac{1}{\tau_{\fvec Q}}+\frac{1}{\tau_{0}^{p}}\right)+VN_{f}\int_{0}^{\Lambda}\frac{\bar{\Delta}_{\omega}\rmd\omega}{\sqrt{\omega^{2}+\bar{\Delta}_{\omega}^{2}}}
\end{align}

From the last equation, it is clear that only the inter-hot-spot non-magnetic
impurity and the intra-hot-spot paramagnetic impurity suppress the
SC gap. The last equation can be solved self-consistently to find
$\bar{\Delta}_{\omega}$; it is convenient then to define an effective
SC order parameter given by:
\[
\Delta=VN_{f}\int_{0}^{\Lambda}\frac{\bar{\Delta}_{\omega}\rmd\omega}{\sqrt{\omega^{2}+\bar{\Delta}_{\omega}^{2}}}\ .
\]

This is the quantity plotted in Fig. 4b of the main text.

\begin{figure}[htbp]
\begin{centering}
\includegraphics[width=0.9\columnwidth]{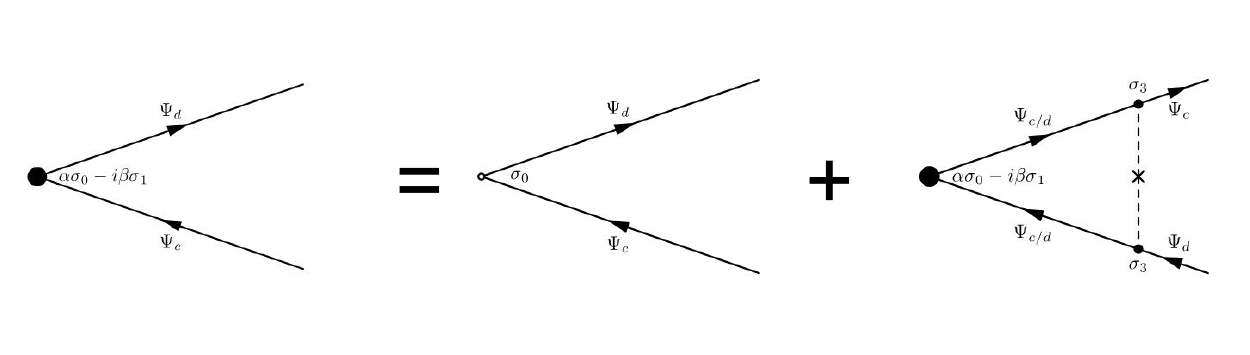}
\par\end{centering}
{\LARGE{}\caption{The spin vertex dressed by the non-magnetic impurity scattering.}
\label{Fig:SpinVertex} }{\LARGE \par}
\end{figure}

 Next, we
consider how the SDW vertex $\Gamma_{\mathrm{SDW}}$ is dressed by
impurity scattering. As shown in Fig.~\ref{Fig:SpinVertex}, the
dressed SDW vertex can be conveniently written in Nambu space as:
\[
\Gamma_{\mathrm{SDW}}=\Gamma_{\mathrm{SDW}}^{\alpha}\left(\Psi_{c}^{\dagger}\sigma_{0}\Psi_{d}+h.c.\right)-i\Gamma_{\mathrm{SDW}}^{\beta}\left(\Psi_{c}^{\dagger}\sigma_{1}\Psi_{d}-h.c.\right)
\]

To calculate the dressed SDW vertex, we need to integrate over the
two dimensional momentum. To calculate these integrals, we apply the
transformation: 
\begin{equation}
\int\rmd^{2}\fvec k\quad\Longrightarrow\quad\int\frac{\rmd\epsilon_{c}\rmd\epsilon_{d}}{|\fvec v_{c}\times\fvec v_{d}|}
\end{equation}

To simply the notation, we define the following quantities (see also Ref. \cite{Chubukov11}): 

\begin{equation}
\begin{split}
F_{\Delta^{2}}= & \frac{\omega(\omega+\Omega)+\bar{\Delta}_{\omega}\bar{\Delta}_{\omega+\Omega}}{\sqrt{\omega^{2}+\bar{\Delta}_{\omega}^{2}}\sqrt{(\omega+\Omega)^{2}+\bar{\Delta}_{\omega+\Omega}^{2}}}\\
F_{\omega\Delta}= & \frac{\omega\bar{\Delta}_{\omega+\Omega}-(\omega+\Omega)\bar{\Delta}_{\omega}}{\sqrt{\omega^{2}+\bar{\Delta}_{\omega}^{2}}\sqrt{(\omega+\Omega)^{2}+\bar{\Delta}_{\omega+\Omega}^{2}}}\\
\tau_{\mathrm{damping}}^{-1}= & N_{f}4|\fvec v_{c}\times\fvec v_{d}|=\frac{\Lambda_{\parallel}v_{f}\sin\theta}{\pi^{2}}
\end{split}
\end{equation}

Direct evaluation of the vertex functions gives the coupled equations:

\begin{equation}
\begin{split}
\Gamma_{\mathrm{SDW}}^{\alpha} & =1+\tau_{\mathrm{damping}}\left[\left(\frac{1}{2\tau_{0}}+\frac{1}{2\tau_{\fvec Q}}\right)-\frac{1}{3}\left(\frac{1}{2\tau_{0}^{p}}+\frac{1}{2\tau_{\fvec Q}^{p}}\right)\right]\left(-\Gamma_{\mathrm{SDW}}^{\alpha}F_{\Delta^{2}}+\Gamma_{\mathrm{SDW}}^{\beta}F_{\omega\Delta}\right)\\
\Gamma_{\mathrm{SDW}}^{\beta} & =\tau_{\mathrm{damping}}\left[\left(\frac{1}{2\tau_{0}}-\frac{1}{2\tau_{\fvec Q}}\right)+\frac{1}{3}\left(\frac{1}{2\tau_{0}^{p}}-\frac{1}{2\tau_{\fvec Q}^{p}}\right)\right]\left(\Gamma_{\mathrm{SDW}}^{\alpha}F_{\omega\Delta}+\Gamma_{\mathrm{SDW}}^{\beta}F_{\Delta^{2}}\right)
\end{split}
\end{equation}

from which we can compute the magnetic susceptibility in Matsubara space:

\begin{equation}
\chi_{\mathrm{SDW}}(\fvec Q,i\Omega)=N_{f}\tau_{\mathrm{damping}}\int\frac{\rmd\omega}{2\pi}\big(\Gamma_{\mathrm{SDW}}^{\alpha}F_{\Delta^{2}}-\Gamma_{\mathrm{SDW}}^{\beta}F_{\omega\Delta}\big)
\end{equation}

Computing the static magnetic susceptibility ($\Omega=0$) gives:
\begin{equation}
F_{\Delta^{2}}=1\ ,F_{\omega\Delta}=0\quad\Longrightarrow\quad\Gamma_{\mathrm{SDW}}^{\alpha}=\left[1+\frac{\tau_{\mathrm{damping}}}{2}\left(\frac{1}{\tau}-\frac{1}{3}\frac{1}{\tau^{p}}\right)\right]^{-1}
\end{equation}
where we defined the total non-magnetic scattering rate $\tau^{-1}=\tau_{0}^{-1}+\tau_{\fvec Q}^{-1}$
and the total paramagnetic scattering rate $\left(\tau^{p}\right)^{-1}=\left(\tau_{0}^{p}\right)^{-1}+\left(\tau_{\fvec Q}^{p}\right)^{-1}$.
Note that the dressed SDW vertex becomes a constant, independent of
the frequency $\omega$. Clearly, while non-magnetic scattering always
reduces $\chi_{\mathrm{SDW}}(\fvec Q,0)$, paramagnetic impurity scattering
enhances it. In Fig. 4 of the main text, we considered the case of
point-like impurities, in which $\tau_{0}^{-1}=\tau_{\fvec Q}^{-1}$
and $\left(\tau_{0}^{p}\right)^{-1}=\left(\tau_{\fvec Q}^{p}\right)^{-1}$.


\begin{thebibliography}{99}
\bibitem{KHnote} We note that an upturn in $C/T$ also has been observed in a commercial Quantum Design PPMS by means of quasiadiabatic thermal relaxation technique.
\end{thebibliography}
\end{document}